\documentstyle[pra,aps,multicol]{revtex}

\def\bxi{{\mbox{\boldmath$\xi$}}}
\def\bR{{\mbox{\boldmath$R$}}}
\def\bS{{\mbox{\boldmath$S$}}}
\def\bE{{\mbox{\boldmath$E$}}}
\def\bF{{\mbox{\boldmath$F$}}}

\def\bD{{\mbox{\boldmath$D$}}}
\def\bC{{\mbox{\boldmath$C$}}}
\def\ba{{\mbox{\boldmath$a$}}}
\def\hba{{\hat{\ba}}}

\def\bq{{\mbox{\boldmath$q$}}}
\def\bk{{\mbox{\boldmath$k$}}}
\def\bh{{\mbox{\boldmath$h$}}}
\def\bz{{\mbox{\boldmath$z$}}}
\def\bv{{\mbox{\boldmath$v$}}}
\def\bzg{{\mbox{\boldmath$\zeta$}}}
\def\bet{{\mbox{\boldmath$\eta$}}}
\def\bxi{{\mbox{\boldmath$\xi$}}}
\def\H{{\cal H}}
\def\L{{\cal L}}
\def\P{{\cal P}}
\def\Z{{\cal Z}}
\def\D{{\cal D}}
\def\Deltlc{\Delta_{{}_0}}

\def\dmu#1{d\mu(#1^*,#1)}
\def\esp#1{e^{\displaystyle #1}}

\def\bave#1{\big\langle#1\big\rangle}
\def\dave#1{\langle\!\langle#1\rangle\!\rangle}
\def\bdave#1{\big\langle\!\!\big\langle#1\big\rangle\!\!\big\rangle}

\def\beq{\begin{equation}}
\def\eeq{\end{equation}}

\def\beqa{\begin{eqnarray}}
\def\eeqa{\end{eqnarray}}

\begin{document}

\title{Effective Hamiltonian with holomorphic variables }

\author{Alessandro Cuccoli}
\address{Dipartimento di Fisica dell'Universit\`a di Firenze
        and Istituto Nazionale di Fisica della Materia (INFM),
        \\ Largo E. Fermi~2, I-50125 Firenze, Italy}

\author{Riccardo Giachetti}
\address{Dipartimento di Fisica dell'Universit\`a di Firenze
        and Istituto Nazionale di Fisica Nucleare (INFN),
        \\ Largo E. Fermi~2, I-50125 Firenze, Italy}

\author{Riccardo Maciocco}
\address{Dipartimento di Fisica dell'Universit\`a di Firenze,
        Istituto Nazionale di Fisica della Materia (INFM),\\
        and Istituto Nazionale di Fisica Nucleare (INFN),
        Largo E. Fermi~2, I-50125 Firenze, Italy}

\author{Valerio Tognetti}
\address{Dipartimento di Fisica dell'Universit\`a di Firenze
        and Istituto Nazionale di Fisica della Materia (INFM),
        \\ Largo E. Fermi~2, I-50125 Firenze, Italy}

\author{Ruggero Vaia}
\address{Istituto di Elettronica Quantistica
        del Consiglio Nazionale delle Ricerche \\ and
        Istituto Nazionale di Fisica della Materia (INFM),
        via Panciatichi~56/30, I-50127 Firenze, Italy}

\date{\today}

\maketitle

\begin{abstract}
The {\it pure-quantum self-consistent harmonic approximation} (PQSCHA)
permits to study a quantum system by means of an effective classical
Hamiltonian -- depending on $\hbar$ and temperature  -- and
classical-like expressions for the averages of observables. In this
work the PQSCHA is derived in terms of the holomorphic variables
connected to a set of bosonic operators. The holomorphic formulation,
based on the path integral for the Weyl symbol of the density matrix,
makes it possible to approach directly general Hamiltonians given in
terms of bosonic creation and annihilation operators.
\end{abstract}

% 05.30.-d Quantum Statistical Mechanics
% 03.65.Sq Quantum Mechanics Semiclassical theories and applications
% 05.30.Jp Boson systems

\begin{multicols}{2}
%=====================================================================

\section {Introduction}
\label{s.i}
The notion of Effective Potential in quantum statistical mechanics was
introduced by Feynman~\cite{FeynmanH65} by means of a variational
method for the path integral with imaginary time. The method was later
on improved by Giachetti and Tognetti~\cite{GT85prl} and Feynman and
Kleinert~\cite{FeynmanK86} in such a way to account completely for the
quantum contribution of the harmonic part of the interaction. The new
formulation, obtained by using the path integral in the Lagrangian
form, has since been successfully applied to several condensed matter
system, usually providing an excellent agreement between theoretical
results and experimental data~\cite{CGTVV95}. The Lagrangian
formulation, however, presents some difficulties for treating systems
for which the kinetic energy and the potential energy are not
separated: spin systems are remarkable instances of such a situation.
In these cases, indeed, the Feynman-Jensen inequality fails to hold
and the variational principle cannot be directly applied. A
generalization that permits to overcome these problems has been found
by using the path integral in the Hamiltonian formulation and the
improved procedure for determining an effective Hamiltonian has been
called the {\it pure-quantum self-consistent harmonic approximation}
(PQSCHA)~\cite{CTVV92ham}. Several applications to condensed matter
have demonstrated its usefulness~\cite{CGTVV95,CTVV96}. Nevertheless,
this generalization can not be applied in a natural way to
field-theory models, which are rather described in terms of creation
and destruction operators: in the bosonic case the natural
classical-like counterpart of these models requires the so called
holomorphic variables $z^*$ and $z$, connected to the creation and
destruction operators.

In order to calculate Hamiltonian path integrals we need a recipe to
associate functions in the phase space to quantum observables~\cite{Louisell73}.
This implies that some care is required to deal with ordering problems: of
course, neglecting the cases where non-local topological terms are
present, the ordering procedure must not affect the final results. It
may however occur that some particular ordering turns out to be more
suited to a given approach or to a specific approximation, in the
sense that it yields more directly the relevant results. The PQSCHA is
a semiclassical expansion and therefore the Weyl ordering with the
associated Wigner distribution~\cite{HilleryCSW84} appears to be the most
convenient choice. Weyl ordering, however, presents also some drawbacks: for
instance it is somewhat cumbersome and its physical meaning is not
immediately evident. Moreover, one can observe that a low-temperature
field theory is rather formulated using the normal ordering, since the
related creation and destruction operators naturally define the vacuum
(possibly close to the ground state) and it allows for an expansion in terms
of Feynman diagrams. On the other hand, the effective Hamiltonian approach
has its main usefulness at intermediate and high temperatures, where
perturbative approaches are useless while quanticity is still significant,
since it fully accounts for the classical nonlinearity of the model under
study, while the quantum character is accounted for at the one-loop level
through suitable renormalization coefficients; furthermore, it can be
verified that, whatever the ordering prescription one starts with,
the final result is such that the effective Hamiltonian, in the limit
of low quantum coupling and/or in the limit of high temperature, reduces
to the Weyl symbol of the original Hamiltonian operator.

In this paper we present the derivation of the effective Hamiltonian
in the framework of PQSCHA for Bose systems in terms of the
holomorphic variables.
In the next Section we present the path integral formulation
for calculating the density matrix, we introduce the Weyl symbols and
their composition properties. We finally establish a relation between
Weyl and normal symbols that turns out to be particularly useful in
the explicit calculation of the former ones.
The third Section shows the formulation of
the PQSCHA in holomorphic variables both in one and in many degrees of
freedom. We introduce the renormalization coefficients that enter the
expressions of the average values of the dynamical variables that, in
turn, determine the effective Hamiltonian. The additional
diagonalization, necessary in the case of many degrees of freedom, is
also given; the same Section presents and uses the low-coupling
approximation, and its specialization to deal with translation invariant
systems, for which much more detailed results can be determined;
eventually, we give an example of the application of the general theory
to a model system with an interaction term expressed by a product of
occupation numbers, thus demonstrating the advantages of the present
formulation.
We finally must add that, for the sake of a more clear exposition, we
have collected in the Appendices the most relevant and less immediate
calculations needed for the development of the subject.

%=====================================================================

\section{Path integral for the density matrix in complex variables}

In order to fix a well defined set of notations, we collect in this
short section some basic properties of the Weyl and normal
representations of operators in terms of complex variables.

Consider a system of $N$ bosons and let $\hba=\{{\hat{a}}_\mu\}_{\mu=1,...,N}$,
${\hba}^{{{\dag}}}=\{{\hat{a}}_\mu^{{\dag}}\}_{\mu=1,...,N}$ be the creation and destruction
operators satisfying the Bose commutation relations. We then introduce the
corresponding classical coordinates
$(\bz^*\!,\bz)\equiv\big\{(z^*_\mu,z_\mu)\big\}_{\mu=1,..,N}$.
An operator ${\hat{a}}t{O}$ can be represented either in the normal ordered or
in the Weyl ordered form, namely
\beqa
 \hat{O}&=&\int\dmu\bk~\esp{i\,{}^t\!{\hba^{{\dag}}}\bk}\esp{i\,{}^t\!\bk^*\hba}
 ~\tilde O_{\scriptscriptstyle\rm{N}} (\bk^*\!,\bk)
\nonumber \\
 &=&\int\dmu\bk~\esp{i({}^t\!{\hba^{{\dag}}}\bk+{}^t\!\bk^*\hba)}
 ~\tilde O(\bk^*\!,\bk) ~.
\label{e.Weylord}
\eeqa
Here the notation ${}^t\!\bq$ is used to denote the transposed matrix
and, for $N$ pairs of complex variables $\bq^*=\{q_\mu^*\}_{\mu=1,...,N}$ and
$\bq=\{q_\mu\}_{\mu=1,...,N}$\,, we have introduced the integration measure
\beq
 \dmu\bq = \prod_\mu {dq_\mu^*dq_\mu\over2\pi i}
 \equiv \prod_\mu {d\Re q_\mu\,d\Im q_\mu\over\pi}~.
\eeq
The normal and Weyl symbols, ${O}_{\scriptscriptstyle\rm N}(\bz^*\!,\bz)$
and ${O}(\bz^*\!,\bz)$, are then the inverse Fourier transforms of
$\tilde{O}_{\scriptscriptstyle\rm N}(\bk^*\!,\bk)$ and
$\tilde{O}(\bk^*\!,\bk)$, according to the general definition
\beq
 f(\bz^*\!,\bz)=\int\dmu\bk~\esp{i({}^t\!\bz^*\bk+{}^t\!\bk^*\bz)}
 ~\tilde{f}(\bk^*\!,\bk) ~.
\eeq
It is easily verified that
\beq
 O(\bz^*\!,\bz)=\esp{-\textstyle{1\over2}\,{}^t\!\partial_{\bz^*}\partial_\bz}
 ~O_{\scriptscriptstyle\rm N}(\bz^*\!,\bz) ~.
\eeq
For practical calculations it is sometimes useful to represent the above
differential operator as follows:
\beq
 \esp{-\textstyle{1\over2}\,{}^t\!\partial_{\bz^*}\partial_\bz}
 \psi(\bz^*\!,\bz)
 =\bdave{\psi(\bz^*{+}\bet^*\!,\bz{-}\bet)}_{\sigma^2={1\over2}} ~,
\label{e.normWeyl1}
\eeq
where $\dave{\,\cdot\,}_{\sigma^2}$ is the Gaussian average
\beq
 \dave{\,\cdot\,}_{\sigma^2} =\sigma^{-2N} \int\dmu\bet~(\,\cdot\,)
 ~\esp{-{}^t\!\bet^*\bet/\sigma^2} \,.
\label{e.davesigma}
\eeq

Weyl symbols have the remarkable property (see Appendix~\ref{a.pi})
\beq
 {\rm{Tr}}\,\big(\hat{O_1}\hat{O}_2\big) =
 \int\dmu\bz~O_1(\bz^*\!,\bz)\,O_2(\bz^*\!,\bz) ~,
\label{e.TrO1O2}
\eeq
that does not extend to products of more than two operators. In particular,
if $\hat\rho=e^{-\beta\hat\H}$ is the density operator at the equilibrium
temperature $T=\beta^{-1}$, we have a classical-like expression of the
statistical averages:
\beq
 \bave{\hat O} = {1\over\Z}~{\rm{Tr}}\,\big(\hat\rho\,\hat O\big)
 = {1\over\Z}~\int\dmu\bz\,\rho(\bz^*\!,\bz)\,O(\bz^*\!,\bz) ~.
\eeq

If, for instance, we consider a single harmonic oscillator with
Hamiltonian $\hat{\H}=\omega({\hat{a}}^{{\dag}}{{\hat{a}}}+{1\over2})$ at inverse
temperature $\beta$ it is straightforward to verify that the Weyl symbol
for $\hat H$ is $\H(z^*\!,z)=\omega{z^*z}$, while for the density matrix,
letting $f=\beta\omega/2$, we first determine the normal symbol
$\rho_{\scriptscriptstyle{\rm{N}}}(z^*\!,z)=\exp[{-f-2z^*z\,\exp(-f)\sinh
f}]$ and, using Eqs.~(\ref{e.normWeyl1}) and~(\ref{e.davesigma}), we
finally get
\beq
 \rho (z^*\!,z) = {1\over\cosh f}\,\esp{-2z^*z\,\tanh f} ~.
\label{e.rhoharmWeyl}
\eeq

Let us finally consider the standard form for the function
$\rho(\bz^*\!,\bz)$ in terms of path-integral:
\beq
 \rho (\bz^*\!,\bz) = \int \D[\bz^*(u),\bz(u)]~\esp{S[\bz^*(u),\bz(u)]} ~,
\label{e.pathint}
\eeq
where $\D[\bz^*(u),\bz(u)]$ is the functional measure and
$S[\bz^*(u),\bz(u)]$ is an Euclidean action, depending on the
functions $\bz^*(u)$ and $\bz(u)$ defined for $0\leq u\leq\beta$. The
external variables $\bz^*$, $\bz$ may explicitly enter the expression
of $S$ through the boundary conditions on the functions $\bz^*(u)$,
$\bz(u)$. In this work, however, we find it convenient to use an
expression of the path-integral in which the boundary variables
$\bz^*(0)$, $\bz(0)$, and $\bz^*(\beta)$, $\bz(\beta)$, are integrated
over and the dependence on $\bz^*$, $\bz$, is explicit in the form of
$S$, namely
\end{multicols}
\noindent\rule{0.5\textwidth}{0.1ex}\rule{0.1ex}{2ex}\hfill

\beqa
 S[\bz^*(u),\bz(u)] &=&
 \int_0^\beta du \bigg\{
 {1\over2}\Big[\,^t\!\dot\bz^*(u)\bz(u)-\,^t\!\bz^*(u)\dot\bz(u)\Big]
 -\H\big(\bz^*(u),\bz(u)\big)\bigg\}
\nonumber \\
 & &~~~~ -{1\over2}\Big\{\,^t\!\bz^*(0)\bz(\beta)-\,^t\!\bz^*(\beta)\bz(0)\Big\}
 - \Big\{\,^t\!\big[\bz^*(\beta)-\bz^*(0)\big]\bz
 -\,^t\!\bz^*\big[\bz(\beta)-\bz(0)\big]\Big\} ~.
\label{e.action}
\eeqa
\hfill\raisebox{-1.9ex}{\rule{0.1ex}{2ex}}\rule{0.5\textwidth}{0.1ex}
\begin{multicols}{2}
For a complete derivation we refer to the exhaustive
analysis of phase space path-integrals by Berezin~\cite{Berezin80}, while a
short but self-contained derivation is presented in Appendix~\ref{a.pi}.

%=====================================================================

\section{PQSCHA: one degree of freedom}

Although the main interest of the method is in the applications to problems
with many degrees of freedom, it is much simpler to understand the method
if we first describe the case of a single degree of freedom.
The idea is to decompose the path-integral expression for $\rho(z^*\!,z)$
into a first sum over all paths sharing the same average point, defined as
the functional
\beq
(\bar{z}^*\!,\bar{z}) =
 {1\over\beta}\int_0^\beta du~\big(z^*(u),z(u)\big)~,
\eeq
and a second sum over average points. In order to do this we introduce
in the path-integral a resolution of the identity that fixes the
average point to $(\bar{z}^*\!,\bar{z})$. It is then natural to define
the reduced density,
\end{multicols}
\noindent\rule{0.5\textwidth}{0.1ex}\rule{0.1ex}{2ex}\hfill

\beq
 \bar\rho(z^*\!,z;\bar{z}^*\!,\bar{z})=\int\D[z^*(u),z(u)]
 ~\delta\bigg(\bar{z}-{1\over\beta}\int_0^\beta du\,z(u)\bigg)
 ~\delta\bigg(\bar{z}^*-{1\over\beta}\int_0^\beta du\,z^*(u)\bigg)
 ~\esp{S[z^*(u),z(u)]} ~,
\label{e.barrho}
\eeq
that collects all contributions coming from paths with the average
point $(\bar{z}^*\!,\bar{z})$, so that the full density reads
\beq
 \rho(z^*\!,z)=\int\dmu{\bar{z}}~\bar\rho(z^*\!,z;\bar{z}^*\!,\bar{z}) ~.
\eeq
We take $\bar\rho(z^*\!,z;\bar{z}^*\!,\bar{z})$ as an unnormalized
probability distribution in the variables $(z^*\!,z)$ and define its
normalization constant as $\exp(-\beta\H_{\rm eff}(\bar{z}^*\!,\bar{z}))$,
so that
\beq
\bar\rho(z^*\!,z;\bar{z}^*\!,\bar{z}) =
 \esp{-\beta\H_{\rm eff}(\bar{z}^*\!,\bar{z})}
 ~\P(z^*\!,z;\bar{z}^*\!,\bar{z}) ~.
\label{e.exHeff}
\eeq
The thermal average of an observable $\hat{O}$ can then be written
\beq
 \bave{\hat O} = {1\over\Z}\int\dmu{\bar{z}} \bigg[\int\dmu{z}~O(z^*\!,z)
 ~\P(z^*\!,z;\bar{z}^*\!,\bar{z})\bigg]
 \esp{-\beta\H_{\rm eff}(\bar{z}^*\!,\bar{z})} ~,
\label{e.classsep}
\eeq
\hfill\raisebox{-1.9ex}{\rule{0.1ex}{2ex}}\rule{0.5\textwidth}{0.1ex}
\begin{multicols}{2}
and it is natural to interpret $\exp[-\beta\H_{\rm eff}(\bar{z}^*\!,\bar{z})]$
as a classical-like effective density, whereas the probability
distribution $P(z^*\!,z;\bar{z}^*\!,\bar{z})$ describes the additional
fluctuations around the point $(\bar{z}^*\!,\bar{z})$. In the classical
limit it can be seen that
$P(z^*\!,z;\bar{z}^*\!,\bar{z})\to\delta\big((z^*\!,z)-(\bar{z}^*\!,\bar{z})\big)$
and $\exp[-\beta\H_{\rm eff}(\bar{z}^*\!,\bar{z})]$ tends to the classical
Boltzmann factor; it follows that the probability $\P$ describes the
pure-quantum fluctuations of the particle, thus providing a separation
between classical-like and pure-quantum contribution to
$\bave{\hat{O}}$. Observe that Eq.~(\ref{e.classsep}) is exact and
provides an ideal starting point for approximations preserving the
full classical nonlinear contribution.

The explicit evaluation of the reduced density
$\bar{\rho}(z^*\!,z;\bar{z}^*\!,\bar{z})$ will be done in a self-consistent
approximation replacing $\H\big(z^*(u),z(u)\big)$ in the action~(\ref{e.action})
with a trial Hamiltonian quadratic in the displacements
from the average point, but depending upon the average points
themselves. This means that the corresponding trial action results
to be non local and therefore cannot be obtained starting from any
quantum operator: we are thus dealing with a larger class of actions.
The parameters of $\H_0$ are optimized independently for any value of
$(\bar{z}^*\!,\bar{z})$. Explicitly we take
\beqa
 \H_0(z^*\!,z;\bar{z}^*\!,\bar{z})
 &=&w(\bar{z}^*\!,\bar{z})+ E(\bar{z}^*\!,\bar{z})(z^*\!-\bar{z}^*)(z-\bar{z})
\nonumber\\
 & &\hspace{9mm}
 +\textstyle{1\over2} \big[ F(\bar{z}^*\!,\bar{z})(z-\bar{z})^2+{\rm c.c.}\big]\,,
\label{e.trialH}
\eeqa
where $E(\bar{z}^*\!,\bar{z})$, $F(\bar{z}^*\!,\bar{z})$ and
$w(\bar{z}^*\!,\bar{z})$ are functions to be optimized. Linear terms
have obviously been neglected since they do not contribute to $S_0$.
It is also evident that a quadratic $\H_0$ yields a Gaussian
probability distribution $\P_0$ and it turns out that it is
centered at the average point. We
therefore introduce the fluctuation variables
$(\xi^*\!,\xi)\equiv(z^*{-}\bar{z}^*\!,z{-}\bar{z})$ and we shall
denote by double brackets the averages over $\P_0$:
\beq
 \dave{\,\cdot\,} = \int\dmu{\xi}~(\,\cdot\,)
 ~\P_0(\bar{z}^*{+}\xi^*\!,\bar{z}{+}\xi;\bar{z}^*\!,\bar{z}) ~.
\eeq
Observe, in particular, that such averages can be easily evaluated
when the moments $\dave{\xi^*\xi}$, $\dave{\xi^2}$ and
$\dave{\xi^{*2}}$ are known.

As shown in Appendix~\ref{a.pqscha}, the explicit result for
$\bar{\rho}_0 (z^*\!,z;\bar{z}^*\!,\bar{z})$ turns out to be
\beq
\bar{\rho}_0 (z^*\!,z;\bar{z}^*\!,\bar{z}) = { f \over {\sinh f}}\,
\esp{- \beta w}\, {2 \over \L(f)}\, \esp{ - 2\,\tilde{\xi}^*
\tilde{\xi}/\L(f) }\,.
\label{e.barrho1}
\eeq
Here $(\tilde{\xi}^*\!,\tilde{\xi})$ are connected to $(\xi^*\!,\xi)$ by a
Bogoliubov transformation,
\beq
\xi^* = R^* \tilde{\xi}^* + S \tilde{\xi}~,~~~~~
\xi = S^* \tilde{\xi}^* + R \tilde{\xi} ~,
\eeq
that diagonalizes the quadratic term,
$E\xi^*\xi+{1\over2}(F\xi^2+{\rm c.c.})=\omega\tilde\xi^*\tilde\xi$\,.
$\L (f)$ is the Langevin function
\beq
 \L(f)=\coth f-1/f ~,
\eeq
where again $f=\beta\omega/2$ while $\omega$ is given by
\beq
 \omega^2=\omega^2(\bar{z}^*\!,\bar{z})
 =E^2(\bar{z}^*\!,\bar{z}) - |F(\bar{z}^*\!,\bar{z})|^2 ~.
\label{e.om1dof}
\eeq
Since no ambiguity can anymore arise, in the following we shall
suppress the bar over $\bar{z}^*$ and $\bar{z}$.

The optimization of the parameters $w(z^*\!,z)$, $E(z^*\!,z)$, and
$F(z^*\!,z)$ is now in order. According to the PQSCHA method we
require that $\P_0$ gives equal averages for the original and the
trial Hamiltonian, as well as for their second
derivatives~\cite{CGTVV95}, namely
\end{multicols}
\noindent\rule{0.5\textwidth}{0.1ex}\rule{0.1ex}{2ex}\hfill

\beq
 \bdave{\H(z^*{+}\xi^*\!,z{+}\xi)} = \bdave{\H_0(z^*{+}\xi^*\!,z{+}\xi)}
 = w(z^*\!,z)+{\textstyle{1\over2}}\,\omega(z^*\!,z)\,\L\big(f(z^*\!,z)\big) ~,
\label{e.daveH1dof}
\eeq
\beqa
 \bdave{\partial_{z^*}\partial_z \H(z^*{+}\xi^*\!,z{+}\xi)} &=&
 \bdave{\partial_{z^*} \partial_z \H_0(z^*{+}\xi^*\!,z{+}\xi)}
 = E(z^*\!,z)
\nonumber \\
 \bdave{\partial_z^2\H(z^*{+}\xi^*\!,z{+}\xi)} &=&
 \bdave{\partial_z^2\H_0(z^*{+}\xi^*\!,z{+}\xi)} = F(z^*\!,z)
\nonumber \\
 \bdave{\partial_{z^*}^2 \H(z^*{+}\xi^*\!,z{+}\xi)} &=&
 \bdave{\partial_{z^*}^2 \H_0(z^*{+}\xi^*\!,z{+}\xi)} = F^*(z^*\!,z)~.
\label{e.sc1dof}
\eeqa
\hfill\raisebox{-1.9ex}{\rule{0.1ex}{2ex}}\rule{0.5\textwidth}{0.1ex}
\begin{multicols}{2}
Comparing Eqs.~(\ref{e.barrho1}) and~(\ref{e.exHeff}), the PQSCHA
effective Hamiltonian can be written as
\beq
 \H_{\rm eff}(z^*\!,z) = w(z^*\!,z)
 +{1\over\beta}\ln\,{\sinh f(z^*\!,z)\over f(z^*\!,z)}~,
\label{e.Heff1dof}
\eeq
and eventually Eq.~(\ref{e.classsep}) for thermal averages is approximated by
\beq
 \bave{\hat{O}} = {1\over\Z}\int\dmu{z}~\bdave{O(z^*{+}\xi^*\!,z{+}\xi)}
 ~\esp{-\beta\H_{\rm{eff}}(z^*\!,z)} ~.
\label{e.aveO}
\eeq

In terms of the transformed variables the second moments of $\P_0$, as
defined in Eq.~(\ref{e.exHeff}), are
$\bdave{\tilde\xi^*\tilde\xi}=\L(f)/2$,
$\bdave{\tilde\xi^*\tilde\xi^*}=\bdave{\tilde\xi\tilde\xi}=0$\,, so that
in the original ones their expressions turn out to be
\beqa
 D(z^*\!,z) = \dave{\xi^*\xi} &=& ~~E ~{\L(f)\over2\omega} ~,
\nonumber \\
 C(z^*\!,z) = ~\dave{\xi^2} &=& -F^* ~{\L(f)\over2\omega} ~,
\nonumber \\
 C^*(z^*\!,z) = \dave{\xi^{*2}} &=& -F ~{\L(f)\over2\omega} ~,
\label{e.ren1dof}
\eeqa
that we call {\it renormalization coefficients}. By means of them
it is useful to introduce a differential operator
$\Delta$ as follows. Letting
\beqa
 \Delta(v,v^*) &=& D(v^*\!,v)\,\partial_{z^*}\partial_z
\nonumber\\
 & &\hspace{5mm}
 +\textstyle{1\over2}
 \,[\,C(v^*\!,v)\,\partial_z^2+C^*(v^*\!,v)\,\partial_{z^*}^2\,]\,.
\label{e.Delta}
\eeqa
we define the action of any function $F(\Delta)$ of the operator
$\Delta$ on a function $a(z^*\!,z)$ as
\beq
 F(\Delta)\,a(z^*\!,z)
 =F\big(\Delta(v^*\!,v)\big)\,a(z^*\!,z)\,\Big|_{(v^*\!,v)=(z^*\!,z)} ~.
\label{e.actDelta}
\eeq
We can then give a useful representation of the double-bracket average
$\dave{\,\cdot\,}$. Indeed, the Gaussian smearing of a generic
function $O(z^*\!,z)$ can be represented as
\beq
 \bdave{O(z^*{+}\xi^*\!,z{+}\xi)} = \esp{\Delta}~O(z^*\!,z)\,.
\label{e.espdelta}
\eeq
For practical calculations, the use of the right-hand-side of
(\ref{e.espdelta}) is particularly convenient when $O(z^*,z)$ is a
low-degree polynomial.

The function $w(z^*\!,z)$ is determined by Eq.~(\ref{e.daveH1dof}),
whose last term can be expressed as
\beq
 \textstyle{1\over2}\,\omega(z^*\!,z)\L\big(f(z^*\!,z)\big)
=\left(\esp\Delta\,\Delta\right)\H(z^*\!,z) ~,
\label{e.oL2del}
\eeq
so
that the effective Hamiltonian can be written in a form with a more
evident renormalization contribution
\beq
 \H_{\rm{eff}}(z^*\!,z)=\Big[\big(1{-}\Delta\big)\,\esp{\Delta}\Big]\,\H(z^*\!,z)
 +{1\over\beta}\ln{\sinh f(z^*\!,z)\over f(z^*\!,z)}~.
\label{e.Heff1dofDelta}
\eeq
It appears therefore that, besides the logarithmic term,
$\H_{\rm{eff}}(z^*\!,z)$ is given by the Weyl Hamiltonian corrected by
terms of the order of the square of the renormalization coefficients.
Moreover, as we shall show in the next Section, the form
(\ref{e.Heff1dofDelta}) is an ideal starting point for a further
approximation in order to deal with many degrees of freedom.

%=====================================================================

\section{PQSCHA: many degrees of freedom}

In this section we are concerned with the generalization of the method
to a system with $N$ degrees of freedom. The self-consistent solution of the
counterpart of Eqs.~(\ref{e.sc1dof})and~(\ref{e.ren1dof}) for
arbitrary values of $(\bar\bz^*\!,\bar\bz)$ is obviously rather
difficult: a further simplification is therefore in order. This is
called the {\em low-coupling approximation} (LCA) and consists in
expanding the parameters --- which are the now $N{\times}N$ complex
matrices $\bE(\bar\bz^*\!,\bar\bz)$, $\bF(\bar\bz^*\!,\bar\bz)$, and
$\bF^*(\bar\bz^*\!,\bar\bz)$ --- so to make the renormalization
coefficients, and hence also the Gaussian averages $\dave{\,\cdot\,}$,
independent of the configuration. Eventually, we describe this scheme
of approximation and we focus, in particular, on translation invariant
systems.

Let us consider therefore a system of $N$ bosons. Again the evaluation of
$\bar\rho(\bz^*\!,\bz;\bar\bz^*\!,\bar\bz)$ is done by replacing
$\H$ with a trial Hamiltonian $\H_0$,
\end{multicols}
\noindent\rule{0.5\textwidth}{0.1ex}\rule{0.1ex}{2ex}\hfill

\beq
 \H_0(\bz^*\!,\bz;\bar\bz^*\!,\bar\bz) = w(\bar\bz^*\!,\bar\bz) +
 ^t\!(\bz^*{-}\bar\bz^*)\,\bE(\bar\bz^*\!,\bar\bz)\,(\bz{-}\bar{\bz})
 +\textstyle{1\over2}\big[\,^t\!(\bz{-}\bar\bz)\,\bF(\bar\bz^*\!,\bar\bz)
 (\bz{-}\bar\bz)+{\rm{c.c.}} \big] ~,
\label{e.H0mdof}
\eeq
\hfill\raisebox{-1.9ex}{\rule{0.1ex}{2ex}}\rule{0.5\textwidth}{0.1ex}
\begin{multicols}{2}
quadratic in the displacements from the average point
\beq
 (\bar\bz^*\!,\bar\bz)={1\over\beta}\int_0^\beta du~\big(\bz^*(u),\bz(u)\big)~.
\label{e.avepoint}
\eeq
As already mentioned, $\bE$, $\bF$, and $\bF^*$ are a Hermitean and
two symmetric $N{\times}N$ complex matrices, respectively.

In expressing $\bar\rho_0$ we again introduce the pure-quantum
fluctuation variables
$(\bxi^*\!,\bxi)=(\bz^*{-}\bar\bz^*\!,\bz{-}\bar\bz)$ and the linear
canonical transformation
\beqa
 \xi_\mu^* &=& \sum_k\big(R_{\mu k}^*(\bar\bz^*\!,\bar\bz)\,\tilde\xi_k^*+
 S_{\mu k}(\bar\bz^*\!,\bar\bz)\,\tilde\xi_k \big)
\nonumber \\
 \xi_\mu &=& \sum_k\big(S_{\mu k}^*(\bar\bz^*\!,\bar\bz)\,\tilde\xi_k^*+
 R_{\mu k}(\bar\bz^*\!,\bar\bz)\,\tilde\xi_k \big)
 \label{e.trandof}
\eeqa
to new variables $(\tilde\bxi^*\!,\tilde\bxi)$ diagonalizes
the quadratic term
\beqa
 & &\sum_{\mu\nu}\Big[ \xi_\mu^*\,E_{\mu\nu}\,\xi_\nu +
 {1\over2}\big(\xi_\mu\,F_{\mu\nu}\,\xi_\nu+{\rm{c.c.}}\big)
 \Big] =
\nonumber\\
 & &\hspace{40mm}
 =\sum_k~\omega_k(\bar\bz^*\!,\bar\bz)~\tilde\xi_k^*\,\tilde\xi_k ~.
\label{e.diagmdof}
\eeqa
Here and in the subsequent discussion transformed variables are
labelled by Latin indices $(k,l,...)$; Greek indices $(\mu,\nu,...)$
are used for the original ones. The diagonalization~(\ref{e.diagmdof}),
that is not possible in the most general case, can
be performed under suitable constraints on the matrices $\bE$, $\bF$,
and $\bF^*$, (see {\it e.g.}~\cite{Colpa78}). However, this problem is
not extremely relevant in the present context, since these matrices
are not external data but parameters that must be optimized: this
constraint just restricts the number of independent matrix elements.
In addition, in most applications the canonical
transformation~(\ref{e.trandof}) is mainly determined by
symmetry considerations.

In analogy to what we did in the previous section, there is no ambiguity
in suppressing the bar over $(\bar{\bz}^*\!,\bar{\bz})$.
The explicit calculation of $\bar\rho_0$, given in Appendix~\ref{a.pqscha},
leads to the following expression for $\H_{\rm{eff}}$:
\beq
 \H_{\rm eff}(\bz^*\!,\bz) = w(\bz^*\!,\bz)
 + {1\over\beta}\sum_k\ln\,{\sinh f_k(\bz^*\!,\bz)\over f_k(\bz^*\!,\bz)}~,
\label{e.Heffmdof}
\eeq
with $f_k(\bz^*\!,\bz)=\beta\omega_k(\bz^*\!,\bz)/2$, while the moments
of $\P_0$ are
$\dave{\tilde\xi_k^*\tilde\xi_l}=\delta_{kl}~\L\big(f_k(\bz^*\!,\bz)\big)/2$
and $\dave{\tilde\xi_k^*\tilde\xi_l^*}=\dave{\tilde\xi_k\tilde\xi_l}=0$\,.
The PQSCHA conditions that determine the optimized parameters $w(\bz^*\!,\bz)$,
$\bE(\bz^*\!,\bz)$, $\bF(\bz^*\!,\bz)$, and $\bF^*(\bz^*\!,\bz)$, are
\end{multicols}
\noindent\rule{0.5\textwidth}{0.1ex}\rule{0.1ex}{2ex}\hfill

\beq
 \bdave{\H(\bz^*{+}\xi^*\!,\bz{+}\xi) } =
 \bdave{\H_0(\bz^*{+}\xi^*\!,\bz{+}\xi) }
 = w(\bz^*\!,\bz) + {1\over2}~ {\sum}_k\,\omega_k(\bz^*\!,\bz)
 ~\L\big(f_k(\bz^*\!,\bz)\big) ~,
\label{e.daveHmdof}
\eeq
\beqa
 \bdave{\partial_{z_\mu^*}\partial_{z_\nu}\H(\bz^*{+}\xi^*\!,\bz{+}\xi)}
 &=& \bdave{\partial_{z_\mu^*}\partial_{z_\nu}\H_0(\bz^*{+}\xi^*\!,\bz{+}\xi)}
 = E_{\mu\nu}(\bz^*\!,\bz)
\nonumber \\
 \bdave{\partial_{z_\mu}\partial_{z_\nu}\H(\bz^*{+}\xi^*\!,\bz{+}\xi)}
 &=& \bdave{\partial_{z_\mu}\partial_{z_\nu}\H_0(\bz^*{+}\xi^*\!,\bz{+}\xi)}
 = F_{\mu\nu}(\bz^*\!,\bz)
\nonumber \\
 \bdave{\partial_{z_\mu^*}\partial_{z_\nu^*}\H(\bz^*{+}\xi^*\!,\bz{+}\xi)}
 &=& \bdave{\partial_{z_\mu^*}\partial_{z_\nu^*}\H_0(\bz^*{+}\xi^*\!,\bz{+}\xi)}
 = {F_{\mu\nu}}^*(\bz^*\!,\bz) ~.
\label{e.scmdof}
\eeqa

Generalizing the procedure of the previous section for the case of one
degree of freedom, we define the renormalization coefficients
\beqa
 D_{\mu\nu}(\bz^*\!,\bz) &=& \dave{\xi_\mu^*\xi_\nu}
 = {1\over2}{\sum}_k\big(R_{\mu k}^*R_{\nu k}+S_{\nu k}^*S_{\mu k}\big)~\L(f_k)
\nonumber \\
 C_{\mu\nu}(\bz^*\!,\bz) &=& \dave{\xi_\mu\xi_\nu}
 = {1\over2}{\sum}_k\big(R_{\mu k}S_{\nu k}^*+R_{\nu k}S_{\mu k}^*\big)~\L(f_k)
\nonumber \\
 C_{\mu\nu}^*(\bz^*\!,\bz) &=& \dave{\xi_\mu^*\xi_\nu^*}
 = {1\over2}{\sum}_k\big(R_{\mu k}^*S_{\nu k}+R_{\nu k}^*S_{\mu k}\big)~\L(f_k)~,
\label{e.defDC}
\eeqa
and, starting from
\beq
 \Delta(\bv^*\!,\bv) = \sum_{\mu\nu} \Big[
 D_{\mu\nu}(\bv^*\!,\bv)\partial_{z_\mu^*}\partial_{z_\nu}
 + {1\over2} \big(C_{\mu\nu}(\bv^*\!,\bv)\partial_{z_\mu}\partial_{z_\nu}
 + C_{\mu\nu}^*(\bv^*\!,\bv)\partial_{z_\mu^*}\partial_{z_\nu^*}\big) \Big]
\eeq
and defining the action of operator $\Delta$ in analogy with
Eq.~(\ref{e.actDelta}), we find that the Gaussian smearing by double-bracket
average can be put into the form
\beq
 \bdave{O(\bz^*{+}\xi^*\!,\bz{+}\xi)} = \esp{\Delta} ~O(\bz^*\!,\bz)\,.
 \eeq
As in the preceding section, it can be easily seen that the effective
Hamiltonian reads
\beq
 \H_{\rm eff}(\bz^*\!,\bz) = \Big[\big(\,1-\Delta\,\big)
 ~\esp{\Delta}\Big] ~\H(\bz^*\!,\bz)\,
 + {1\over\beta}\sum_k\ln\,{\sinh f_k(\bz^*\!,\bz)\over f_k(\bz^*\!,\bz)}\,,
\label{e.HeffmdofDelta}
\eeq
and the general expression for calculating thermal averages takes the form:
\beq
 \bave{\hat{O}} = {1\over\Z}\int\dmu{\bz}~\bdave{O(\bz^*{+}\bxi^*\!,\bz{+}\bxi)}
 ~\esp{-\beta\H_{\rm{eff}}(\bz^*\!,\bz)} ~.
\label{e.aveOmdof}
\eeq
\hfill\raisebox{-1.9ex}{\rule{0.1ex}{2ex}}\rule{0.5\textwidth}{0.1ex}
\begin{multicols}{2}

%=====================================================================
\medskip

We now describe the low-coupling approximation, whose main purpose is
to make the averages $\dave{\,\cdot\,}$ independent of the
configuration~\cite{CGTVV95}.
The self-consistent equations need therefore to be
solved only once, with a great simplification for implementing the
method. One can think of several ways to obtain this goal:
their choice substantially depends on the physics of the problem~\cite{CTVV92ham}.
The simplest way, however, consists in expanding the matrices $\bE(\bz^*\!,\bz)$,
$\bF(\bz^*\!,\bz)$ around a
self-consistent minimum $(\bz_0^* ,\bz_0)$ of $\H_{\rm eff}$:
\beqa
 \bE(\bz^*\!,\bz) &=& \bE + \delta \bE(\bz^*\!,\bz)
\nonumber \\
 \bF(\bz^*\!,\bz) &=& \bF + \delta \bF(\bz^*\!,\bz) ~,
\eeqa
where $\bE= \bE(\bz_0^*,\bz_0)$, $\bF=\bF(\bz_0^*,\bz_0)$, thus using the
convention to drop the arguments of functions
evaluated at $(\bz_0^*,\bz_0)$.
As a consequence the frequencies will be expanded to the first order,
$\omega_k(\bz^*\!,\bz)= \omega_{k} + \delta \omega_k(\bz^*\!,\bz)$ and
for the evaluation of the averages we shall only take $\bD$ and $\bC$ as
renormalization coefficients. It is explicitly shown in Appendix~\ref{a.LCA}
that the effective Hamiltonian becomes
\beqa
 \H_{\rm eff}(\bz^*\!,\bz) &=& \esp{\Deltlc}\, \H(\bz^*\!,\bz)
 -\Deltlc\,\esp{\Deltlc}\,\H(\bz_0^*,\bz_0)
\nonumber\\
 & & \hspace{23mm}
 +{1\over\beta}\sum_k\ln\,{\sinh f_k\over f_k } ~,
\label{e.HeffLCA}
\eeqa
where $f_k=\beta\omega_k/2$ and
\beq
 \Deltlc = \sum_{\mu\nu} \left[
 D_{\mu \nu} \partial_{z_\mu^*}\partial_{z_\nu}
 +{1\over2} \left( C_{\mu\nu}\partial_{z_\mu}\partial_{z_\nu}
 +C_{\mu\nu}^*\partial_{z_\mu^*}\partial_{z_\nu^*} \right) \right] \,.
\eeq
The general formula for thermal averages~(\ref{e.aveOmdof}) still holds
provided that the Gaussian smearing of $O(\bz^*\!,\bz)$ is calculated with the
LCA renormalization coefficients, {\it i.e.} as $\exp(\Deltlc)\,O(\bz^*\!,\bz)$.

It often occurs that the indices $\mu,\nu,\dots$ refer to the sites of
a lattice, whose symmetries can sometimes be very helpful in order to
simplify the analysis, provided that the minimum configuration of
$\H_{\rm eff}(\bz^*\!,\bz)$ shares the same property. The calculations
are very easy in the case of translation symmetry. In particular, for
a one-dimensional lattice
\beq
 E_{\mu\nu} = E_{\mu-\nu}, ~~~~ F_{\mu\nu} = F_{\mu-\nu}\,,
\eeq
where the hermiticity of $\bE$ and the symmetry of $\bF$
imply that the components $E_\alpha$ and $F_\alpha$ satisfy
\beq
 E_\alpha = {E_{-\alpha}}^*,~~~~ F_\alpha = F_{- \alpha} ~.
\label{e.erefsim}
\eeq
Performing a Fourier transformation,
\beq
 \xi_\mu = N^{- {1 \over 2}} \sum_k { \xi_k \esp{ i k \mu} },~~
 \xi_\mu^* = N^{- {1 \over 2}} \sum_k { \xi_k^* \esp{ - i k \mu} } ~,
\label{e.fourier}
\eeq
the left-hand-side of~(\ref{e.diagmdof}) becomes
\beq
 \sum_k{ \left[ E_k \xi_k^* \xi_k + {1 \over 2} ( F_k \xi_k
 \xi_{-k} + {\rm c.c.} ) \right] }=\sum_k{ \,\omega_k\,
 \tilde{\xi}^*_k\tilde{\xi}_k} ~,
\label{e.tihk}
\eeq
where
\beq
 E_k = \sum_\alpha E_\alpha \esp{-ik\alpha}
 ~,~~~~ F_k = \sum_\alpha F_\alpha \esp{-ik\alpha} ~.
\eeq
As a consequence of~(\ref{e.erefsim}) $E_k$ is real and $F_k =
F_{-k}$. If the lattice has more than one dimension the generalization
is obvious and, in particular, the summation over wave-vectors is
performed over the first Brillouin zone. If there are internal degrees
of freedom for each lattice site, the Fourier transform must be
followed by a diagonalization in the internal space. Finally, the
diagonalization of the quadratic form~(\ref{e.tihk}) is completed by a
Bogoliubov transformation
\beqa
 \xi_k^* &=& A_k {\tilde{\xi}_k}^* - B_k \tilde{\xi}_{-k} \nonumber \\
 \xi_{-k} &=& - {B_{-k}}^* {\tilde{\xi}_{k}}^* + {A_{-k}}^* \tilde{\xi}_{-k} ~,
\label{e.bog}
\eeqa
where the canonical conditions imply
\beq
 A_k^* {A_k} - B_k^* B_k = 1, ~~~ A_k = A_{-k}, ~~~ B_k = B_{-k} ~,
\eeq
and it turns out that
\beq
 \omega_k ( A_k^* A_k + B_k^* B_k) = E_k ~,~~~~
 \omega_k A_k B_k = {F_k}/2 ~.
\eeq
Finally,
\beq
 \omega_k^2 = {E_k}^2 - {F_k}^* F_k ~.
\eeq
By combining the transformations~(\ref{e.fourier} )and~(\ref{e.bog})
we obtain the LCA approximation of the matrices $\bR$ and $\bS$
defined in~(\ref{e.trandof}):
\beq
 R_{\mu k} = N^{-{1 \over 2}} A_k^* \esp{i k \mu} ,~~~~~
 S_{\mu k}^* = - N^{-{1 \over 2}} B_k^* \esp{- i k \mu}
\eeq
Using then~(\ref{e.defDC}) we
also can specify the LCA renormalization coefficients:
\beqa
 D_{\mu\nu}&=& {1\over N}\sum_k E_k\,{\L(f_k)\over2\omega_k}\cos k(\mu{-}\nu)~,
\nonumber \\
 C_{\mu\nu}&=&-{1\over N}\sum_k F_k^*\,{\L(f_k)\over2\omega_k}\cos k(\mu{-}\nu)~,
\nonumber \\
 C_{\mu\nu}^*&=&-{1\over N}\sum_k F_k\,{\L(f_k)\over2\omega_k}\cos k(\mu{-}\nu)~.
\label{e.dfunef}
\eeqa
In many cases only a reduced
set of these coefficients will be explicitly needed in $\H_{\rm eff}$.
For instance, any model with on-site nonlinearity entails only on-site
coefficients $D=D_{\mu\mu}$, $C=C_{\mu \mu}$ which are independent of $\mu$
due to translation invariance.

%=====================================================================
\section{An application}

As an example of a system that is more conveniently treated by means
of holomorphic variables, we  shall consider a model Hamiltonian with
quartic on-site interaction,
\beq
\hat{\H} = -\sum_{\bf i\,d} \,{\hat{a}}_{\bf i}^{{\dag}}\,
 {\hat{a}}_{{\bf i}+{\bf d}}\,
+ \sum_{\bf i} \Big( V_1\, \hat{n}_{\bf i} + V_2\, \hat{n}_{\bf i}^2 \Big) ~,
\label{e.boseh}
\eeq
where ${\bf i}$ runs over sites of a Bravais lattice, ${\bf d}$
denotes the displacements towards the $Z$ nearest-neighbour,
$\hat{n}_{\bf i}={\hat{a}}_{\bf i}^{{\dag}}{\hat{a}}_{\bf i}$,
and $V_1$ and $V_2>0$ are constants.
One can recognize in (\ref{e.boseh})
the Hamiltonian of the well-known Bose-Hubbard model
\cite{FisherWGF89,KashurnikovS96}. Its Weyl symbol is easily found to be
\beqa
 \H(\bz^*\!,\bz) &=& - {N \over 2} V_1 -
 \sum_{\bf i\,d} \,z_{\bf i}^*\, z_{{\bf i}+{\bf d}}\,
\nonumber\\
 & & \hspace{5mm}
 +\sum_{\bf i} \Big[\,(V_1-V_2) z_{\bf i}^* z_{\bf i}
 + V_2 (z_{\bf i}^* z_{\bf i})^2 \Big] ~.
\eeqa
$\H$ is invariant under global phase transformations, and provided
that the LCA effective Hamiltonian shares the same property, the
translation invariant minimum $(z_{0{\bf i}}^*,z_{0{\bf i}})=(z_0^*,z_0)$
of $\H_{\rm{eff}}$ is degenerate; therefore we can
choose as representative the real minimum with $z_0=z_0^*$. It is easy
to verify that equations~(\ref{e.scmdof}) for the matrices $\bE$ and
$\bF$ become
\beqa
 E_{\bf ij}
 &=& \big[ \, (V_1-V_2) + 4V_2(z_0^2 + D)\, \big] \delta_{\bf ij}
 - \sum_{\bf d}\,\delta_{{\bf i},{\bf j}+{\bf d}}  ~,
\nonumber \\
 F_{\bf ij} &=& 2V_2( z_0^2 + C)\, \delta_{\bf ij}~,
\eeqa
and $F^*_{\bf ij}=F_{\bf ij}$. They yield the
following Fourier transforms:
\beqa
 E_{\bf k} &=& (V_1-V_2) + 4V_2(z_0^2 + D)
 - \sum_{\bf d}\, \cos ({\bf k}{\cdot}{\bf d}) ~,
\nonumber \\
 F_{\bf k} &=& F = 2V_2(z_0^2 + C) ~,
\eeqa
The renormalization coefficients read now
\beqa
 D&=& {1\over N}\sum_{\bf k} E_{\bf k}\,
 {\L(f_{\bf k})\over2\omega_{\bf k}}~,
\nonumber \\
 C&=&-{F\over N}\sum_{\bf k} \,{\L(f_{\bf k})\over2\omega_{\bf k}}~,
\eeqa
with $\omega^2_{\bf k}=E_{\bf k}^2-F^2$.

Using~(\ref{e.HeffLCA}), the effective Hamiltonian can be eventually written as
\beqa
 \H_{\rm eff}(\bz^*\!,\bz) &=& {\cal{G}}(\beta) + \H(\bz^*\!,\bz)
\nonumber\\
 & &\hspace{2mm}
 + V_2 \sum_{\bf i} \left[ 4 D ~ z_{\bf i}^* z_{\bf i}
 + C ({z_{\bf i}^*}^2 + z_{\bf i}^2) \right] ~,
\eeqa
where
\beqa
 {\cal{G}}(\beta) &=& - V_2N \Big[ 2z_0^2(2 D + C) + 2D^2 + C^2 \Big]
\nonumber\\
 & &\hspace{28mm}
 + {1\over\beta} \sum_{\bf k}\ln\,{\sinh f_{\bf k}\over f_{\bf k}}
\eeqa
is a uniform contribution that affects the partition function but
cancels out in the expression of thermal averages. The (real)
translation-invariant stationary points of $\H_{\rm eff}$ are given by
\beqa
 z_{0,1} &=& 0
\nonumber\\
 z_{0,2}^2 &=& {V_2-V_1+Z\over2V_2}-2D-C ~.
\eeqa
One can check that when $z_{0,2}^2<0$ the minimum is $z_{0,1}$
and in the above formulas one should replace $z_0=z_{0,1}$;
conversely, for $z_{0,2}^2>0$ the minimum is $z_{0,2}$.
Since the calculation of the renormalization coefficients depends on
the chosen minimum, both cases can simultaneously occur: if this is
the case, the relevant minimum to be used at a given temperature is
the one which gives the smaller free energy.

%=====================================================================

\section{Summary}

We have presented a construction of the effective Hamiltonian suitable
for the study of the thermodynamics of field-theory models, with
Hamiltonian expressed in terms of creation and  destruction operators.
The construction is done by a path-integral approach, in terms of the
holomorphic variables $z^*$ and $z$. The main result we presented
consists in Eq.~(\ref{e.aveO}) for one degree of freedom, and in
Eq.~(\ref{e.aveOmdof}) for many degrees of freedom: in the latter case
the LCA effective Hamiltonian~(\ref{e.HeffLCA}) allows one to perform
practical calculations. As in the previous formulation~\cite{CGTVV95}
in terms of phase-space conjugate variables, $p$ and $q$, the
framework retains all classical nonlinear effects embodied in the
classical-like formulas for the thermal averages and the partition
function. The quantum effects related to the quadratic part of the
Hamiltonian are completely accounted for, while the pure-quantum
nonlinearity is treated  at one-loop (Hartree-Fock) level. The
low-coupling approximation permits to actually deal with many degrees
of freedom, so that our results can be used for a direct application
to finite-temperature field theory. We finally observe that this
approach can be extended to deal with the thermodynamics of Fermi
systems \cite{CuccoliGMTV99}.

\end{multicols}
\noindent\rule{0.5\textwidth}{0.1ex}\rule{0.1ex}{2ex}\hfill

%=====================================================================
%======================= Appendices ==================================
%=====================================================================

\appendix

%=====================================================================

\section{Path-integral for $\rho$ in terms of complex variables}
\label{a.pi}

We give here a short derivation of the expression for $\rho(\bz^*\!,\bz)$
as a path-integral. Although the subject is rather well-known, we have chosen
to include it since we want to stress those aspects connected with the
Weyl symbols. We start from the Trotter's formula
for a discrete imaginary time $u\to\varepsilon{i}$ ($i=1,...,M$,
$\varepsilon\equiv M/\beta$) and we take then the continuum limit:
\beq
 \hat{\rho}=\esp{-\beta\hat\H}
 =\lim_{M\to\infty}\big(1-\varepsilon\hat\H\big)^M~.
\label{e.Trotter}
\eeq

In order to introduce the Weyl formulation, it is necessary to write the
expression of the Weyl symbol of the product of $M$ operators.
Let us first derive the product rule for two operators $\hat{O}$ and
$\hat{O}_1$, i.e. the Weyl symbol $O{*}O_1(\bz^*\!,\bz)$
for $\hat{O}\hat{O}_1$\,. Using Eq.~(\ref{e.Weylord}) and the
Baker-Campbell-Hausdorff identity, we have
\beq
 \hat{O} \hat{O}_1 = \int \dmu{\bk_0} \int \dmu{\bk_1}
 ~\esp{i[{}^t\!{\hba^{{\dag}}}(\bk_0{+}\bk_1)+{}^t\!(\bk_0^*{+}\bk_1^*)\hba]}
 ~\esp{\textstyle{1\over2}({}^t\!{\bk_1^*}\bk_0{-}{}^t\!\bk_0^*\bk_1)}
 \tilde O(\bk_0^*,\bk_0)\tilde{O}_{1}(\bk_1^*,\bk_1) ~,
\eeq
Using now the expression of the inverse Fourier transform
and the representation of the $\delta$-function
\beq
 \delta(\bk^*)\delta(\bk)=\int\dmu\bz
 ~\esp{({}^t\!{\bz^*}\bk+{}^t\!\bk^*\bz)} ~,
\label{e.delta1}
\eeq
we can determine the product rule for the Weyl symbols that is traditionally
called {\it star product}:
\beqa
 O{*}O_1(\bz^*\!,\bz) &=&
 \int\dmu{\bh_1}\,\dmu{\bk_1}
 ~\esp{i[\,^t\!\bz^*(\bh_1{+}\bk_1)+\,^t\!(\bh_1^*{+}\bk_1^*)\bz]}
 ~\esp{{1\over2}(\,^t\!\bk_1^*\bh_1-\,^t\!\bh_1^*\bk_1)}
\nonumber \\
 & & \times\int\dmu{\bv_1}\,\dmu{\bz_1}
 ~\esp{-i(\,^t\!\bv_1^*\bh_1{+}\,^t\!\bh_1^*\bv_1)}
 ~\esp{-i(\,^t\!\bz_1^*\bk_1{+}\,^t\!\bk_1^*\bz_1)}
 ~O(\bv_1^*,\bv_1)\,O_1(\bz_1^*,\bz_1) ~.
\nonumber \\
 &=& 2^{2N}\int\dmu{\bv_1}\,\dmu{\bz_1}
 ~O(\bv_1^*,\bv_1)\,O_1(\bz_1^*,\bz_1)
 ~\esp{-2[\,^t\!(\bv_1^*{-}\bz^*)(\bv_1{-}\bz_1)
 -\,^t\!(\bv_1^*{-}\bz_1^*)(\bv_1-\bz)] }~.
\label{e.Weylprod2}
\eeqa
The integration of this expression, using~(\ref{e.delta1}),
immediately yields Eq.~(\ref{e.TrO1O2}).
By iterating Eq.~(\ref{e.Weylprod2}), we obtain the star product of
three operators $\hat{O}$, $\hat{O}_1$, $\hat{O}_2$:
\beqa
 O{*}O_1{*}O_2(\bz^*\!,\bz) &=&
 2^{4N}\int\dmu{\bv_1}\,\dmu{\bz_1}\,\dmu{\bv_2}\,\dmu{\bz_2}\,
 O (\bv_1^*,\bv_1)\,O_1 (\bz_1^*,\bz_1)\,O_2 (\bz_2^*,\bz_2)
\nonumber \\
 & & \times \esp{-2[\,^t\!(\bv_1^*{-}\bv_2^*)(\bv_1{-}\bz_1)
 -\,^t\!(\bv_1^*{-}\bz_1^*)(\bv_1{-}\bv_2)]}
 ~\esp{-2[\,^t\!(\bv_2^*{-}\bz^*)(\bv_2{-}\bz_2)
 -\,^t\!(\bv_2^*{-}\bz_2^*)(\bv_2{-}\bz)]} ~.
\eeqa
It is now easy to see that in general, if we take $\hat{O}=\hat{1}$
and therefore $O=1$, we can write
\begin{equation}
 \mathop{*}_{i=1}^M O_i(\bz^*\!,\bz)
 = 2^{2MN} \!\!\int\!\bigg[\prod_{i=1}^M
 \dmu{\bv_i}\,\dmu{\bz_i}\,O_i(\bz_i^*,\bz_i) \bigg]
 \exp\bigg\{\!\!-2\sum_{i=1}^M
 \big[\,^t\!(\bv_i^*{-}\bv_{i+1}^*)(\bv_i{-}\bz_i)
 {-}\,^t\!(\bv_i^*{-}\bz_i^*)(\bv_i{-}\bv_{i+1})\big] \!\bigg\} ,
\end{equation}
where $\bv_{M+1}\equiv\bz$. Letting $\bv_i=\bz_i{+}\bzg_i$ (with
$\bzg_{M+1}=\bz$) and $\Delta\bz_i=\bz_i{-}\bz_{i-1}$, the previous
expression becomes
\begin{equation}
 \mathop{*}_{i=1}^M O_i(\bz^*\!,\bz)
 = 2^{2MN} \!\!\int\bigg[\prod_{i=1}^M
 \dmu{\bzg_i}\,\dmu{\bz_i}\,O_i(\bz_i^*,\bz_i) \bigg]
 \exp\bigg\{2\sum_{i=1}^M
 \big[\,^t\!(\bzg_{i+1}^*{+}\Delta\bz_{i+1}^*)\bzg_i
 -\,^t\!\bzg_i^*(\bzg_{i+1}{+}\Delta\bz_{i+1})\big] \bigg\} ~.
\end{equation}
The integrations over the variables $\bzg_i$ are performed taking into account
a different representation of the $\delta$-function, namely
\beq
 \delta(\bk)\delta(\bk^*)=\int\dmu\bz~\esp{\pm(\,^t\!\bz^*\bk-\,^t\!\bk^*\bz)}~,
\label{e.delta2}
\eeq
Since no generality is lost in assuming $M$ to be an even number, we
eventually obtain
\beqa
 \mathop{*}_{i=1}^M O_i(\bz^*\!,\bz)
 &=& 2^{MN} \!\!\int\bigg[\prod_{i=1}^M\dmu{\bz_i}\,O_i(\bz_i^*,\bz_i) \bigg]
\nonumber \\
 & &~~~~~~~~ \times
 \exp\bigg\{ 2\bigg[ -\sum_{k=1}^{M/2}\sum_{j=1}^k
 \big(\,^t\!\Delta\bz_{2k+1}^*\Delta\bz_{2j}
 -\,^t\!\Delta\bz_{2j}^*\Delta\bz_{2k+1}\big)
 +\sum_{j=1}^{M/2}\big(\,^t\!\Delta\bz_{2j}^*\,\bz
 -\,^t\!\bz^*\Delta\bz_{2j}\big) \bigg]
 \bigg\} ~.
\label{e.Weylprodn}
\eeqa
Finally, according to Eq.~(\ref{e.Trotter}), we set
$O_i=1-\varepsilon\H\simeq{e}^{-\varepsilon\H}$,
and in the continuum limit the integrand becomes
\beqa
 &-& \int_0^\beta du ~\H\big(\bz^*(u),\bz(u)\big)
 - 2 \int_0^{\beta/2}\!\!\! du \int_0^u \! du'
 \big[\,^t\!\dot\bz^*(2u)\,\dot\bz(2u')-\,^t\!\dot\bz^*(2u')\,\dot\bz(2u) \big]
\nonumber \\
 & & + 2\int_0^{\beta/2}\!\!\! du'
 \big[\,^t\!\bz^*(\beta)\,\dot\bz(2u')-\,^t\!\dot\bz^*(2u')\,\bz(\beta) \big]
 +2 \int_0^{\beta/2}\!\!\! du'
 \big[\,^t\!\dot\bz^*(2u')\,\bz-\,^t\!\bz^*\,\dot\bz(2u')\big] ~.
\eeqa
The integration over $u'$ is trivial and, by the change of variable
$u\to\beta{-}u$, we obtain the path integral given by Eqs.~(\ref{e.pathint})
and~(\ref{e.action}), with the measure $\D[\bz^*(u),\bz(u)]$ formally
defined by
\beq
 \D[\bz^*(u),\bz(u)]=\lim_{M\to\infty}2^{MN}\prod_{i=1}^M\dmu{\bz_i}~.
\eeq

%=====================================================================

\section{The PQSCHA method}
\label{a.pqscha}

Let us evaluate the reduced density $\bar\rho_0$~(\ref{e.barrho})
with the trial Hamiltonian $\H_0$, Eq.~(\ref{e.trialH}).
Representing the $\delta$-function as in Eq.~(\ref{e.delta2}) one gets
\beq
 \bar\rho_0(z^*\!,z;\bar{z}^*\!,\bar{z})
 = \int\dmu{v} \esp{\bar{z}^*v{-}v^*\bar{z}} \int \D\big[z^*(u),z(u)\big]
 ~\esp{S_0[z^*(u),z(u)]-{\textstyle{1\over\beta}\int_0^\beta}
 du [z^*(u)\,v-v^*z(u)] } ~,
\eeq
where $S_0$ is given by Eq.~(\ref{e.action}) with $\H_0$.
We change the integration variables in the path integral to
$\big(\xi^*(u),\xi(u)\big)=\big(z^*(u){-}\bar{z}^*\!,z(u){-}\bar{z}\big)$
and we make the Bogoliubov transformation
\beq
 \xi^*(u) =
 R^*\tilde\xi^*(u)+S\,\tilde\xi(u),~~~\xi(u)= S^*\tilde\xi^*(u)+R\,\tilde\xi(u) ~,
\eeq
with $R^*R-S^*S=1$. This permits to diagonalize the quadratic term
\beq
 E\,\xi^*(u)\,\xi(u)+\textstyle{1\over2}\big[F\xi^2(u)+{\rm{c.c.}}\big]
 = \omega~\tilde\xi^*(u)\,\tilde\xi(u) ~.
\eeq
It is straightforward to verify that this transformation preserves at the
same time the functional measure and the form of the exponent in the path
integral. Performing the same transformation on $(z^*\!,z)$,
$(\bar{z}^*\!,\bar{z})$ and $(v^*\!,v)$ we obtain
\beqa
 \bar\rho_0(z^*\!,z;\bar{z}^*\!,\bar{z}) &=& \esp{-\beta w}
 \int\dmu{\tilde{v}} \int\D\big[\tilde\xi^*(u),\tilde{\xi}(u)\big]
\nonumber \\
 & &~~~~ \times \exp\bigg\{ \int_0^\beta du \bigg[
 -{1\over\beta}\Big[\tilde\xi^*(u)\,\tilde{v}{-}\tilde{v}^*\tilde\xi(u)\Big]
 +{1\over2}\Big[\dot{\tilde\xi}^*(u)\,\tilde\xi(u){-}
 \tilde\xi^*(u)\,\dot{\tilde\xi}(u) \Big]
 -\omega\,\tilde\xi^*(u)\,\tilde\xi(u) \bigg] +
\nonumber \\
 & &~~~~~~~~~~~~ - {1\over2} \Big[\tilde\xi^*(0)\,\tilde\xi(\beta)
 {-}\tilde\xi^*(\beta)\,\tilde\xi(0) \Big]
 -\Big[\big[\tilde\xi^*(\beta){-}\tilde\xi^*(0)\big] (\tilde{z}{-}\tilde{\bar{z}})
 -(\tilde{z}^*{-}\tilde{\bar{z}}^*)\big[\tilde\xi(\beta){-}\tilde\xi(0)\big]\Big]
 \bigg\} ~.
\eeqa
The shift
\beq
 \big(\xi^*(u),\xi(u)\big) \longrightarrow
 \Big(\xi^*(u)+{\tilde{v}^*\over\beta\omega}\, ,\,
 \xi(u)-{\tilde{v}\over\beta\omega}\Big)
\eeq
eliminates the linear term, and the outcoming expression of
$\bar\rho_0$ contains the path integral of a single harmonic oscillator:
\beqa
 \bar\rho_0(z^*\!,z;\bar{z}^*\!,\bar{z}) &=& \esp{-\beta w}
 \int\dmu{\tilde{v}} ~\esp{-{\tilde{v}^*\tilde{v}/\beta\omega}}
 \int\D\big[\tilde\xi^*(u),\tilde\xi(u)\big]
\nonumber \\
 & &~~~~ \times \exp\bigg\{ \int_0^\beta du \bigg[
 {1\over2} \Big[\dot{\tilde\xi}^*(u)\,\tilde\xi(u)-
 \tilde\xi^*(u)\,\dot{\tilde\xi}(u) \Big]
 -\omega\,\tilde\xi^*(u)\,\tilde\xi(u)\bigg]
 -{1\over2}\Big[\tilde\xi^*(0)\,\tilde\xi(\beta)
 -\tilde\xi^*(\beta)\,\tilde{\xi}(0)\Big]
\nonumber \\
 & &~~~~~~~~~~~~ - \bigg[\big[\tilde\xi^*(\beta)-\tilde\xi^*(0)\big]
 \Big(\tilde{z}-\tilde{\bar{z}}+{\tilde{v}\over\beta\omega}\Big)
 -\Big(\tilde{z}^*-\tilde{\bar{z}}^*-{\tilde{v}^*\over\beta\omega}\Big)
 \big[ \tilde\xi(\beta)-\tilde\xi(0)\big] \bigg] \bigg\} ~.
\eeqa
Therefore, using Eq.~(\ref{e.rhoharmWeyl}) we obtain
\beqa
 \bar\rho_0(z^*\!,z;\bar{z}^*\!,\bar{z}) &=& \esp{-\beta w}
 \int\dmu{\tilde{v}} ~\esp{-{\tilde{v}^*\tilde{v}/\beta\omega}}
 {1\over\cosh f}~\exp\Big\{
 -2\Big(\tilde{z}^*{-}\tilde{\bar{z}}^*{-}{\tilde{v}^*\over\beta\omega}\Big)
 \Big(\tilde{z}{-}\tilde{\bar{z}}{+}{\tilde{v}\over\beta\omega}\Big) \tanh f
 \Big\} ~,
\eeqa
where $f=\beta\omega/2$. Finally, performing the integration over
$(\tilde{v}^*\!,\tilde{v})$, we arrive at Eq.~(\ref{e.barrho1}).

Eq.~(\ref{e.Heff1dofDelta}) can be derived from Eq.~(\ref{e.oL2del}); the
latter, in turn, can be easily verified in this way:
\beq
 \esp{\Delta}\Delta\H(z^*\!,z)=\bdave{\Delta\H(z^*{+}\xi^*\!,z{+}\xi)}
 =D(z^*\!,z)\bdave{\partial_{z^*}\partial_{z}\H(z^*{+}\xi^*\!,z{+}\xi)}
 + {\textstyle{1\over2}}
 \big[C(z^*\!,z)\bdave{\partial_{z}^{2}\H(z^*{+}\xi^*\!,z{+}\xi)}+{\rm c.c.}
 \big] \,.
\eeq
Using Eq.~(\ref{e.sc1dof}) and Eq.~(\ref{e.ren1dof}) the right hand
side becomes
\beq
 D(z^*\!,z)E(z^*\!,z)+{\textstyle{1\over2}}\big[C(z^*\!,z)F(z^*\!,z)+{\rm c.c.}\big] =
 \big[E^2(z^*\!,z)-F^*(z^*\!,z)F(z^*\!,z)\big]
 {{\L(f(z^*\!,z))}\over{2\omega(z^*\!,z)}} ~,
\eeq
and finally, Eq.~(\ref{e.oL2del}) follows from Eq.~(\ref{e.om1dof}).

In order to deal with the case of many degrees of freedom we use the
representation~(\ref{e.delta2}) of the $\delta$-function for
implementing the constraint~(\ref{e.avepoint}), so that the expression
for $\bar\rho_0(\bz^*\!,\bz;\bar\bz^*\!,\bar\bz)$ becomes
\beq
 \bar\rho_0(\bz^*\!,\bz;\bar\bz^*\!,\bar\bz) =
 \int\!\dmu{\bv}~ e^{^t\!\bar\bz^*\bv-\,^t\!\bv^*\bar\bz}
 \int\! \D\big[\bz^*(u),\bz(u)\big]
 ~\exp\bigg\{S_0\big[\bz^*(u),\bz(u)\big]
 -{1\over\beta}\int_0^\beta \!\!du
 \big[\,^t\!\bz^*(u)\bv{-}\,^t\!\bv^*\bz(u)\big] \bigg\} \,,
\label{e.rhomdof}
\eeq
where $S_0$ is the action~(\ref{e.action}) with the trial
Hamiltonian~(\ref{e.H0mdof}). This expression is evaluated by making
the linear canonical transformation~(\ref{e.trandof}) which permits to
decouple the path integral into a product of one dimensional harmonic
path integrals. The same transformation, done over all the pairs of
conjugate variables that appear in~(\ref{e.rhomdof}), yields the form
$\bar\rho_0=\exp(-\beta w)\prod_k (\bar\rho_0)_k$ for the reduced
density. Taking into account~(\ref{e.diagmdof}) we can use the result
for one degree of freedom Eq.~(\ref{e.barrho1}) and  we obtain
\beq
 \bar\rho_0(\bz^*\!,\bz;\bar{\bz}^*\!,\bar{\bz}) = \esp{- \beta w}
 \prod_k \left( { f_k \over {\sinh f_k}} {2 \over \L(f_k)}
 \exp\Big[ - {2 \over \L(f_k)} \tilde{\xi}_k^* \tilde{\xi}_k \Big] \right) ~,
\eeq
where $f_k=\beta\omega_k/2$.

%=====================================================================
\section{Low-coupling approximation}
\label{a.LCA}
We give here the steps which lead to the LCA expression~(\ref{e.HeffLCA}) for
$\H_{\rm eff}$, when the LCA is made by
expanding the renormalization coefficients around the self-consistent
minimum $(\bz_0^*,\bz_0)$ of $\H_{\rm eff}$.

We start by expanding Eq.~(\ref{e.HeffmdofDelta}) to the lowest order in
the difference $ \delta \omega_k(\bz^*\!,\bz) = \omega_k(\bz^*\!,\bz)
- \omega_{k}$, recalling the convention to drop the arguments of functions
evaluated at $(\bz_0^*,\bz_0)$. The expansion of the logarithmic term
is
\beqa
 {1\over\beta}\sum_k\ln\,{\sinh f_k(\bz^*\!,\bz)\over f_k(\bz^*\!,\bz)} &=&
 {1\over\beta}\sum_k\ln\,{\sinh f_k\over f_k}
 +\sum_k {\L(f_k)\over2}~\delta\omega_k(\bz^*\!,\bz)
\\ \nonumber
 &=& {1\over\beta}\sum_k\ln\,{\sinh f_k\over f_k}
 + \Big[\esp{\Delta}\Deltlc\H(\bz^*\!,\bz)
 - \esp{\Deltlc}\Deltlc\H(\bz_0,\bz_0^*)\Big]~,
\eeqa
where we have taken into account the fact that Eq.~(\ref{e.oL2del})
generalizes to
\beq
 \sum_k\,\omega_k(\bz^*\!,\bz) \,{\L\big(f_k(\bz^*\!,\bz)\big)\over  2} =
 \Big[\,\esp{\Delta}\Delta\,\Big]\,\H(\bz^*\!,\bz)\,.
\eeq
Letting $\Delta = \Deltlc + \delta\Delta$, we have the expansion
\beq
 \Big[(1 - \Delta) \esp{\Delta}\Big] \H (\bz^*\!,\bz)
 = \Big[\,(1 - \delta \Delta ) \esp{\delta \Delta}\,\Big]
 \esp{\Deltlc} \H (\bz^*\!,\bz)-\esp{\Delta}\Deltlc\H(\bz^*\!,\bz)
 = \esp{\Deltlc} \H (\bz^*\!,\bz)-\esp{\Delta}\Deltlc \H(\bz^*\!,\bz) \,,
\eeq
so that, eventually, Eq.~(\ref{e.HeffLCA}) is obtained.

%====================================================================
%=============  references  =========================================

%\bibliographystyle{prsty}
%\bibliography{ourbib,biblio,bibina}

\end{document}